\documentclass[twocolumn]{aastex631}
\usepackage{graphicx}
\usepackage{float}
\usepackage{rotating}
\usepackage{tablefootnote}
\DeclareUnicodeCharacter{2212}{\ensuremath{-}}

\newcommand{\lsim }{{\lower0.8ex\hbox{$\buildrel <\over\sim$}}}
\newcommand{\gsim }{{\lower0.8ex\hbox{$\buildrel >\over\sim$}}}
\newcommand{\Msun}{\ifmmode {M_{\odot}}\else${M_{\odot}}$\fi}
\newcommand{\Lsun}{\ifmmode {L_{\odot}}\else${L_{\odot}}$\fi}
\newcommand{\Rsun}{\ifmmode {R_{\odot}}\else${R_{\odot}}$\fi}

\shorttitle{Classifying Compact Radio Emission in Nearby Galaxies}
\shortauthors{Dage et al.}

\begin{document}

\title{Classifying Compact Radio Emission in Nearby Galaxies: a 10GHz Study of Active Galactic Nuclei, Supernovae, Anomalous Microwave Emission and Star Forming Regions}

\correspondingauthor{Kristen Dage}
\email{kristen.dage@curtin.edu.au}

\author[0000-0002-8532-4025]{Kristen C. Dage}
\affiliation{International Centre for Radio Astronomy Research -- Curtin University, GPO Box U1987, Perth, WA 6845, Australia}

\author[0000-0001-8424-2848]{Eric W. Koch}
\affiliation{Center for Astrophysics $|$ Harvard \& Smithsonian, 60 Garden St., Cambridge, MA 02138, US}

\author[0000-0001-8424-2848]{Evangelia Tremou}
\affiliation{National Radio Astronomy Observatory, Socorro, NM 87801, USA}

\author[0000-0003-1814-8620]{Kwangmin Oh}
\affiliation{Center for Data Intensive and Time Domain Astronomy, Department of Physics and Astronomy, Michigan State University, East Lansing, MI 48824, USA \\} 

\author[0000-0001-9261-1738]{Susmita Sett}
\affiliation{International Centre for Radio Astronomy Research -- Curtin University, GPO Box U1987, Perth, WA 6845, Australia}

\author[0000-0002-1185-2810]{Cosima Eibensteiner}

\affiliation{National Radio Astronomy Observatory, 520 Edgemont Road, Charlottesville, VA 22903, USA}
 \altaffiliation{Jansky Fellow of the National Radio Astronomy Observatory}
\author[0000-0002-1000-6081]{Sean T. Linden}
\affiliation{Steward Observatory, University of Arizona, 933 N Cherry Avenue, Tucson, AZ 85721, USA}
\author[0000-0003-0851-7082]{Angiraben D. Mahida}
\affiliation{International Centre for Radio Astronomy Research -- Curtin University, GPO Box U1987, Perth, WA 6845, Australia}

\author[0000-0001-7089-7325]{Eric J. Murphy}
\affiliation{National Radio Astronomy Observatory, 520 Edgemont Road, Charlottesville, VA 22903, USA}

\author[0009-0004-9310-020X]{Muhammad Ridha Aldhalemi}
\affiliation{Henry Ford College, 5101 Evergreen Rd, Dearborn, MI 48128, USA \\} 

\author[0009-0006-8696-9892]{Zainab Bustani}
\affiliation{Henry Ford College, 5101 Evergreen Rd, Dearborn, MI 48128, USA \\} 
\author[0009-0005-2051-1304]{Mariam Ismail Fawaz}
\affiliation{Henry Ford College, 5101 Evergreen Rd, Dearborn, MI 48128, USA \\}

\author[0009-0006-5976-8120]{Hans J. Harff}
\affiliation{Henry Ford College, 5101 Evergreen Rd, Dearborn, MI 48128, USA \\}

\author[0009-0000-6403-8903]{Amna Khalyleh}
\affiliation{Henry Ford College, 5101 Evergreen Rd, Dearborn, MI 48128, USA \\}

\author[0009-0007-9611-1774]{Timothy McBride}
\affiliation{Henry Ford College, 5101 Evergreen Rd, Dearborn, MI 48128, USA \\}

\author[0009-0006-2324-0738]{Jesse Mason}
\affiliation{Henry Ford College, 5101 Evergreen Rd, Dearborn, MI 48128, USA \\}

\author[0009-0000-8689-3476]{Anthony Preston}
\affiliation{Henry Ford College, 5101 Evergreen Rd, Dearborn, MI 48128, USA \\}

\author[0009-0001-7561-6753]{Cortney Rinehart}
\affiliation{Henry Ford College, 5101 Evergreen Rd, Dearborn, MI 48128, USA \\}

\author[0009-0003-1543-4514]{Ethan Vinson}
\affiliation{Henry Ford College, 5101 Evergreen Rd, Dearborn, MI 48128, USA \\} 

\author[0000-0001-8424-2848]{Teresa Panurach}
\affiliation{Center for Materials Research, Department of Physics, Norfolk State University, Norfolk VA 23504, USA \\}

\author[0000-0002-7092-0326]{Richard M. Plotkin}
\affiliation{Department of Physics, University of Nevada, Reno, NV 89557, USA}
\affiliation{Nevada Center for Astrophysics, University of Nevada, Las Vegas, NV 89154, USA}
\author[0000-0002-9396-7215]{Liliana Rivera Sandoval}

\affiliation{Department of Physics and Astronomy, University of Texas Rio Grande Valley, Brownsville, TX 78520, USA}

\begin{abstract}
We present 115 compact radio point sources in three galaxies,  NGC 5474, NGC 4631 and M51, taken in the most extended (A-)configuration of the Karl G. Jansky Very Large Array at 10GHz. Several of these compact radio point sources have diffuse counterparts identified in previous multi-band studies of resolved radio continuum emission. We find compact counterparts to eight star forming regions, four anomalous microwave emission candidates, and one supernova remnant (SN 2011dh). Nine of the compact radio sources match X-ray counterparts, the majority of which are background galaxies. These AGN are all within the D25 (isophotal diameter) of the host galaxy and might act as contaminants for X-ray binary population studies, highlighting the need for high-resolution multi-band imaging. This study showcases the broad number of science cases that require sensitive radio facilities, like the upcoming Square Kilometre Array and the planned next generation Very Large Array.

\end{abstract}


\section{Introduction}  

Multi-wavelength spectral energy distributions (SEDs) are a common approach to physically classify astronomical objects.
Radio emission mechanisms can provide important constraints to enable these classifications, however, its use has largely lagged behind imaging at shorter wavelengths due to the difficulty of combining both high spatial resolution and sensitivity. In radio, we are able to probe very diffuse galaxy emission with the most compact arrays, resolved diffuse emission (e.g. from star formation regions) from less compact arrays, all the way to very compact emission probing jets produced by black holes with the most extended configurations.

At GHz frequencies, we can probe several physical mechanisms like synchrotron and free-free emission. These can provide galaxy-scale links to star formation rate since both supernova remnants  (SNRs) emitting synchrotron (via Type II SNe) and free-free emission from $H_{II}$regions trace different evolutionary points of massive stars.

Furthermore, the excess measured at $\sim$10 of GHz points to another radio emission mechanism, aptly named anomalous microwave emission (AME).
First discovered from Galactic foreground models for cosmic microwave background experiments, these excesses were first found to correlate with mid-IR dust emission \citep{Kogut1996,deOliveira-Costa1997,Leitch1997}.
The current, most widely accepted model for AME is the spinning dust model, where the rapid rotation of very small grains with a non-zero electric dipole moment can generate microwave emission \citep{Draine1998a,Draine1998b,Draine1999}.
\citet{Murphy2010} reported the first detection of likely AME beyond the Milky Way in the active star-forming galaxy NGC 6946, and subsequent $10$s~GHz observations in nearby galaxies, particularly the 33 AME candidates presented in \citet{Linden20}, are beginning to enable population-level studies.
The combination of a larger observed AME population resolved on a wider range of spatial scales will aid in pinpointing the underlying source as spinning dust or another emission mechanism.

Given the broad range of spatial scales that these various radio emission mechanisms can emit on, classifying radio-emitting sources and connecting to the broader electromagnetic spectrum are often limited by the spatial resolution of the observations.
In particular, classifying the radio emission from rarer objects---including compact object and intermediate-mass black holes (IMBHs)---\emph{requires} strong constraints on more numerous galactic (SNRs, $H_{II}$ regions, AME sources) and background (AGN) radio sources, all of which will otherwise act as a contaminant.
Sensitive, resolved radio surveys that at least match the scales traced at shorter wavelengths are crucial to enable these classifications and interpret the unresolved radio emission, particularly for high-z systems.

The largest, uniform and resolved GHz radio survey of nearby galaxies to date is the Star Formation in Radio Survey presented in \citet{Linden20}, where they obtained VLA 3 GHz, 15 GHz and 33 GHz observations of 50 star forming galaxies and classify 335 sources, 238 of which are star-forming regions (SFRs), and 33 of which are AME.

In this paper, we expand on previous surveys to search for compact emission from $\sim0\farcs15$ A-configuration X-band observations captured by the Karl G. Jansky Very Large Array (VLA) \citep{1980ApJS...44..151T} that may be associated with resolved sources classified by the multi-band observations by \cite{Linden20} in three galaxies: M51 (NGC 5194), NGC 4631 and NGC 5474.
These three galaxies have extensive coverage with \textit{Hubble Space Telescope} (HST) from the Legacy ExtraGalactic UV Survey \citep[LEGUS;][]{2015AJ....149...51C} or other archival observations, and have archival \textit{Chandra} X-ray observations.
Where possible, we use the optical and X-ray catalogs to improve constraints on characterizing our X-band VLA sources.
\cite{2012MNRAS.419.2095M} also studied high mass X-ray binary (HMXB) populations in the D25 \citep[isophotal diameter, e.g., ][]{Paturel1987} of these galaxies, so we are also able to provide an assessment of their contamination rate.

The face-on, grand-spiral galaxy M51\citep[$d=8.58$ Mpc; ][]{McQuinn16} is among the best-studied nearby galaxies, and particularly due to its low-luminosity AGN and central radio jet structures, has been extensively studied at radio wavelengths. In addition to sources catalogued by \citet{Linden20}, compact radio emission was previously studied by \cite{Maddox07} (pre-VLA upgrade) who identified a number of supernova remnants. VLBI studies by \cite{Rampadarath} found several supernova remnants and background galaxies, and subarcsecond imaging with LOFAR \citep{Venkattu23} was able to provide constraints on the low frequency properties of many of these systems. 

Our other two targets are similarly nearby and have been extensively studied at shorter wavelengths.
NGC4631 is an edge-on interacting system at $d=7.3$~Mpc \citep{2013AJ....146...86T} and the most massive galaxy in the NGC 4631 group.
This galaxy hosts a central starburst driving a wind characterized via its X-ray emission \citep{Wang1995}.

Its edge on geometry and wind have made this galaxy a key target for studying the diffuse radio emission and magnetic field structure at the disk-halo interface \citep{Mora2013,Wiegert2015}.
Our final target is NGC 5474 (d=6.8 Mpc; \citealt{2015AstL...41..239T}), a peculiar dwarf galaxy in the M101 group whose current disrupted structure likely arises from tidal interactions with M101 \citep{Linden2022}.
Prior L-band radio observations of this galaxy in the \citet{Condon1987} catalog focused on characterizing the diffuse radio component.

   With new A-configuration VLA observations, we reach $<10$~pc scales with sensitive coverage thanks to the VLA upgraded bandwidth. At these scales, we can (i) identify which sources remain compact and unresolved relative to the \citet{Linden20} catalog, and (ii) compare compact radio sources to multi-wavelength studies to improve object classification.
    By comparing to X-ray, we can also assess contamination rates in previous object classification for, e.g., HMXB populations identified in the X-ray.
    
    While this is only a study of a few galaxies, with new facilities like the Square Kilometre Array (SKA) coming online and the upcoming next generation VLA (ngVLA), these results show the power and promise of a highly sensitive radio facility to probe a broad range of science cases. 

\begin{table*}[]
\caption{Galaxy properties and derived parameters from z0MGS \citep{Leroy2019}. SFRs measured from a combination of GALEX-FUV and WISE4, and $\Delta$MS is the offset from the $z=0$ star-forming main sequence \citep[equation 19 from]{Leroy2019}.  Distances from \cite{2013AJ....146...86T} (NGC 4631),  \cite{2015AstL...41..239T} (NGC 5474), \cite{McQuinn16} (M51).}
\label{table:sample}
\begin{tabular}{lllll}
Galaxy & Distance (Mpc) & log (M$_\star$ / M$_\odot$) & log (SFR / M$_\odot$ yr$^{-1}$) & $\Delta$MS \\ \hline \hline
NGC4631 & 7.3 & 10.05 & 0.30 & 0.38 \\ 
M51 & 8.58 & 10.73 & 0.65 & 0.28 \\ 
NGC5474 & 6.8 & 8.67 & -1.27 & -0.28 \\ 
\end{tabular}
\end{table*}
Throughout this paper, we make use of the terms ``diffuse'', ``resolved'', and ``compact'' radio emission with respect to the physical scales, processes, and the angular resolution of the data in consideration.
We reserve ``diffuse'' radio emission to refer to the wide-spread kpc non-thermal radio component \citep{Beck2015}, ``resolved'' to refer to the individual sources within (or in the background of) the galaxies as observed by previous radio surveys at coarser angular resolution \citep[e.g.,][]{Linden20}, and ``compact'' to the point-source or marginally resolved structures recovered in our $0.15\arcsec\sim$~few pc resolution data that we present here.

\begin{table*}[]
\begin{flushleft}
\label{table:n4631}
\caption{Compact emission in NGC 4631 from our X-band observations with multi-band counterparts from the \citet{Linden20} catalog (L20). We infer the diffuse 10GHz flux density from the spectral indices and 15GHz flux density from \cite{Linden20}, and report the ratio of diffuse emission to compact emission in the `Ratio' column. The positional uncertainty in R.A. and Dec is denoted as the last decimal in parentheses.  All flux densities in mJy.}
\begin{tabular}{llllllllll}
Class$_{\rm X}$ & S$_{\rm X}$ (10 GHz)   &  Class$_{\rm L20}$  & S3GHz & S15 GHz & S33 GHz &Ratio & R.A.                    \& Dec               & Sep. ($\arcsec$)\\ \hline \hline
AGN   & 0.45    $\pm$ 0.04 $\times 10^{-1}$ & SFR & 0.06   & 0.03$\times 10^{-2}$   & 3.07 & 66.3  & 12:42:04.29 (4) +32:32:25.41 (2)   & 1.17  \\
(??)  &  0.42   $\pm$ 0.04 $\times 10^{-1}$  & AME 1 & 1.75  & 0.87   & 1.25  &19.3 & 12:42:05.62 (3) +32:32:30.25 (4)   & 0.79 \\
(??)  & 2.06 $\pm$ 0.41 $\times 10^{-1}$  & AME 2 & 1.54  & 0.53   & 0.94  &2.3 & 12:42:06.23 (2)+32:32:31.85(2)& 0.23 \\ \hline
\end{tabular}
\end{flushleft}
\end{table*}

\section{Data and Analysis}
\begin{figure}
\includegraphics[width=0.48\textwidth]{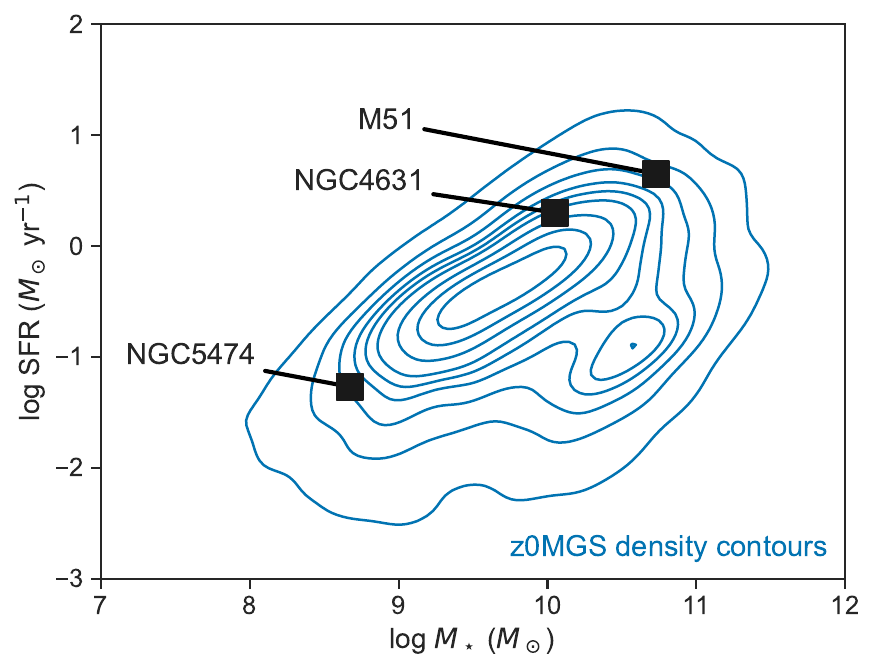}
\caption{Our three targeted galaxies and their locations relative to the $z=0$ massive galaxy populations from z0MGS \citep{Leroy2019}. Though a small sample, we highlight that all of these sources lie close to the locus of star-forming main sequence galaxies.}
\label{Fig:sfms}
\end{figure}

We reduced compact (A-configuration), X-band radio from the VLA and archival X-ray observations from \textit{Chandra} of three galaxies, NGC 4631 (7.3 Mpc), NGC 5474 (6.8 Mpc) and M51 (8.58 Mpc).
Table \ref{table:sample} provides basic properties of our sample.
Figure \ref{Fig:sfms} compares our sample with the nearby galaxy population within the $d<50$~Mpc z0MGS sample based on WISE and GALEX \citep{Leroy2019}.
Though small, all 3 of our targets are close to the star forming main sequence and span the full range of SFR and M$_*$ of the z=0 multi-wavelength Galaxy synthesis (z0MGS) sample, making it a representative pilot study.
We describe the reduction process below.

\subsection{X-ray analysis}

For our analysis, we utilized the {\it Chandra} Interactive Analysis of Observations (CIAO, version 4.15.1) software \citep{Fruscione06}, employing the latest calibration files from the Chandra Calibration Database (CALDB). We reprocessed all the data using the \texttt{chandra\_repro} script, focusing on the full 0.3--10.0 keV energy band to ensure a comprehensive analysis of the X-ray sources.

For M51, we processed the longest available observations from 2012 (PI: Kuntz): ObsIDs 13812 (2012-09-12, 160ks), 13813 (2012-09-09, 180ks), 18314 (2012-09-20, 190ks), 13815 (2012-09-23, 68ks) and 13816 (2012-09-26, 74 ks) observed in 2012 (PI: Kuntz), all ACIS-S. These datasets were merged with \texttt{merge\_obs} tool to maximize the detections of event files and it provides a unified dataset for our flux estimation. X-ray source detection was performed with the \texttt{wavdetect} algorithm, utilizing a range of spatial scales of 2, 4, 8, 16, 24, 32, 48 pixels. A significance threshold of $10^{-6}$ was applied, corresponding to approximately 1 false alarm per $1024 \times 1024$ pixel image. We considered the sources detected with a significance larger than $3\sigma$.

The unabsorbed X-ray fluxes ($f_{x}$) for these sources were computed from \texttt{srcflux} tool within the 0.3--10.0 keV energy range. For this computation, we assumed a hydrogen column density (\texttt{hden}) of $3.8 \times 10^{18} \, \text{cm}^{-2}$ \citep{2023MNRAS.521.2719Y}. The fluxes were calculated based on an absorbed power-law model with a photon index fixed at 1.7. For the X-ray luminosity calculation, we used $L_{x}=4\pi d^{2}f_{x}$ and $d=8.58$ Mpc was used for the distance to M51.

\begin{figure}
 \centering 
 \includegraphics[width=0.46\textwidth]{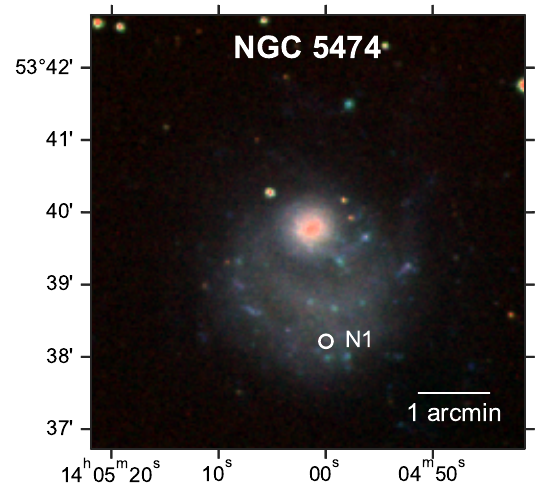}\\
 \caption{As in Figure \ref{fig:ngc5194_rgb_matched_xband} for NGC~5474, with position N1 marking the background galaxy. \label{fig:ngc5474_rgb_matched_xband}}
\end{figure}

\begin{figure}
 \centering 
 \includegraphics[width=0.48\textwidth]{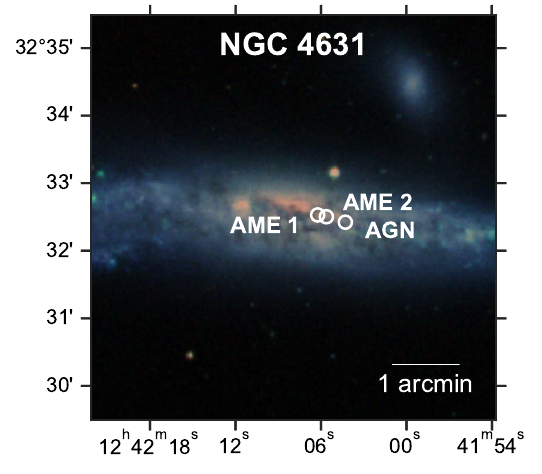}\\
 \caption{As in Figure \ref{fig:ngc5194_rgb_matched_xband} for NGC~4631. Our work identifies the AGN based on the resolved radio lobe morphology (see \S\ref{sub:ngc4631}); this source was previously classified as a SFR in \citet{Linden20} and a HMXB by \citet{2012MNRAS.419.2095M}. \label{fig:ngc4631_rgb_matched_xband}}
\end{figure}

\begin{figure}
 \centering 
 \includegraphics[width=0.46\textwidth]{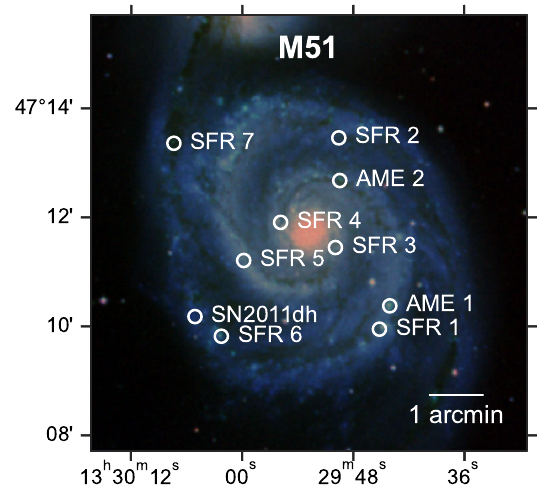}\\
 \caption{Location of X-band radio sources in M51 that match with archival multi-band radio sources \citep{Linden20} or \emph{Chandra} X-ray detections overlaid on DSS2 imaging. The star formation regions (SFRs) and anomalous microwave emission (AME) classifications come from \cite{Linden20}. \label{fig:ngc5194_rgb_matched_xband}}
\end{figure}

\begin{figure}
\includegraphics[width=0.49\textwidth]{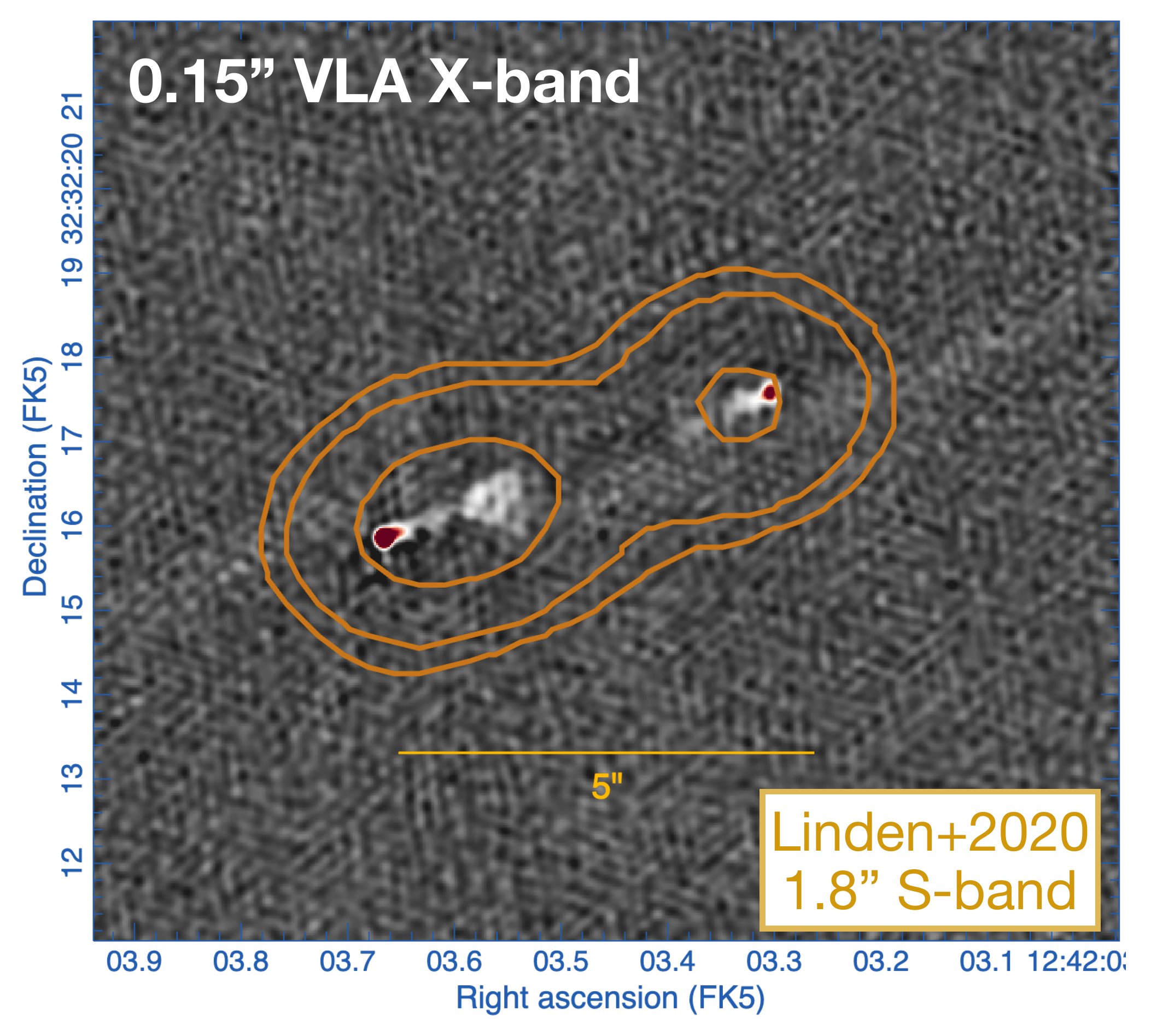}
\caption{Our resolved observations reveal clear AGN jets at position 12:42:03.59 +32:32:16.28 within the NGC~4631 field, strongly suggesting this is a background radio-bright AGN.
However, this source has previously been catalogued as a star-forming region (SFR) based on lower resolution radio SED modeling by \citet{Linden20}, and as a HMXB from the coincident X-ray emission by \citet{2012MNRAS.419.2095M}. The orange band contours show 1.8" diffuse S-band emission at 0.3, 0.6, and 2~mJy \citep{Linden20}. Our finding emphasizes the need for resolved radio surveys to distinguish sources within the host and background objects.
 }
\label{Fig:AGNs}
\end{figure}

NGC 5474 and NGC 4631 have not been targeted by \textit{Chandra} as frequently as M51. For each, we use the longest available observation.  For NGC 5474, we used ObsID 9546 (PI: Jenkins), a 30ks ACIS-S observation taken on 2007-12-03. 
For NGC 4361, we use a 60ks ACIS-S observation, ObsID 797 (PI: Wang), taken on 2000-04-16. We reduce these with the same method as described above, assuming fixed line-of-sight neutral hydrogen densities of 1.16 and 1.29 $\times 10^{20} \, \text{cm}^{-2}$ respectively from \textit{Chandra}'s \textsc{colden} \footnote{\url{https://cxc.harvard.edu/toolkit/colden.jsp}}.

\subsection{Radio Data}
The field of NGC 5474, and four separate fields of M51 were observed by the VLA in the most extended A-array configuration between June-July 2023 (NRAO/VLA Program ID 23A-104, PI: K.Dage), with X-band receivers (8-12 GHz),  3-bit samplers, with two independent 2048-MHz wide basebands centered at 9.0 GHz and 11.0 GHz. The bandwidth was divided into 128-MHz wide spectral windows, and each spectral window was sampled by 64 channels. Phase calibrators were J1335+4542 for the M51 fields and J1419+5423 for the NGC 5474 field. The bandpass and flux calibrator was 3C286 (1331+305).
The total time of each observing block was 2 hours long with approximately 1.1 hours on source. 

For NGC 4631, we used archival VLA data taken with VLA A-configuration using the X-band receiver and the same instrument setup as for the M51 and NGC 5474. Three observations from the NRAO/VLA Program ID 22A-068, PI: J. Mangum were used. All three observations were taken between May - June 2022. NGC 4631 field was targeted for 1 hour total per observing epoch (3 hours total). The quasar J1310+3220 was used as flux and phase calibrator.

We followed standard reduction procedures with the Common Astronomy Software Application \cite[CASA;][]{2007casa,2022casa} to calibrate the data and produce images, using the 
Briggs weighting scheme \cite[robust=0,][]{1995briggs} and frequency dependent clean components (with two Taylor terms; nterms=2) to mitigate large-bandwidth effects \citep{2012rau}. For each primary beam corrected image, the mean RMS noise was 4~$\mu$Jy/beam with a median synthesized beam in the images of 0.16" $\times$ 0.14".

We searched for point sources detected to a 5 $\sigma$ threshold using \textsc{PyBDSF} \citep{2015ascl.soft02007M}. For one of the sources in NGC 4631 (AME 2), \textsc{PyBDSF} split it into two different sources. We instead report the flux from a single source as measured by Carta, assuming 20\% uncertainty. 

We cross-matched the compact radio both with the X-ray point source catalog, and also with the published classified diffuse emission catalog from \cite{Linden20}. For both cases, we did this using \textsc{topcat}. We used a 1" matching tolerance between X-ray to radio (\textit{Chandra}'s nominal positional uncertainty), and 2.5" between compact and diffuse emission, which is the nominal size of the diffuse regions in \cite{Linden20}. These classifications are displayed in Figures \ref{fig:ngc5474_rgb_matched_xband} (NGC 5474), \ref{fig:ngc4631_rgb_matched_xband} (NGC 4631) and \ref{fig:ngc5194_rgb_matched_xband} (M51).

\begin{table*}[]
\begin{flushleft}
    
\caption{Compact emission (A-configuration) in M51 with resolved B-configuration counterparts detected in \citet{Linden20}. We again present the ratio of inferred 10 GHz diffuse flux density to measured 10GHz compact flux density in the `Ratio' column. The positional uncertainty in R.A. and Dec is denoted as the last decimal in parentheses. All flux densities in mJy.}
\label{table:m51linden}
\begin{tabular}{lllllllllll}
Class$_{\rm L20}$ & S3GHz    & S15GHz & S33GHz  & Ratio  & R.A.                 \&   Dec.                 &  S$_{\rm X}$ (10 GHz)    & Sep. \\ \hline \hline
AME  1  & 1.33     & 0.46   & 0.58    &6.4 & 13:29:44.06 (3)+47:10:22.73 (6) & 6.13 $\pm$ 0.82 $\times 10^{-2}$       & 0.35 \\
SFR  1 & 0.81     & 0.55   & 0.58    &12.1 & 13:29:45.18 (4) +47:09:57.06 (8) & 4.32  $\pm$ 0.81 $\times 10^{-2}$       & 1.84 \\
AME  2 & 0.54     & 0.46   & 0.72  &13.0   & 13:29:49.46 (6) +47:12:40.75 (4) & 3.66 $\pm$ 0.67 $\times 10^{-2}$       & 0.75 \\
SFR  2 & $<$ 0.13 & 0.1    & 0.15  &3.4   & 13:29:49.56 (5) +47:13:27.75 (5) &  3.57 $\pm$ 0.71 $\times 10^{-2}$       & 0.30 \\
SFR  3 & 2.92     & 1.22   & 0.29   &19.3  & 13:29:49.93 (5) +47:11:26.98 (5) & 5.02 $\pm$  0.82 $\times 10^{-2}$       & 1.13 \\
SFR  4 & 0.71     & 0.28   & 0.16  &3.4   & 13:29:55.85 (3) +47:11:54.67 (2)  & 6.52 $\pm$ 0.72 $\times 10^{-2}$       & 1.23 \\
SFR  5 & 0.4      & 0.36   & 0.38    &2.8 & 13:29:59.84 (2) +47:11:12.66 (3)  & 1.28 $\pm$ 0.07 $\times 10^{-1}$       & 0.06 \\
SFR  6 & 1.2      & 0.4    & 0.34   &8.1  & 13:30:02.25 (4) +47:09:49.26 (3) & 3.86 $\pm$ 0.65 $\times 10^{-2}$       & 1.52 \\
SNe   (SN 2011dh) & 1.54     & 0.17   & $<$ 0.15 &1.9& 13:30:05.10 (4) +47:10:10.91 (3) & 5.07 $\pm$ 0.74 $\times 10^{-2}$       & 0.16 \\
SFR  7 & 0.29     & 0.2    & $<$ 0.98 & 37.1 &13:30:07.39 (9) +47:13:21.75 (8) & 2.92 $\pm$ 0.49 $\times 10^{-2}$       & 0.31 \\ \hline
\end{tabular}
\end{flushleft}

\end{table*}

\section{Results and Discussion} \label{sec:results}
We were able to detect 115 compact radio point sources across all three galaxies. We cross matched these to existing studies of resolved radio emission from \cite{Linden20}, as well as X-ray from \textit{Chandra}, and compared to previous radio studies in other bands/resolution. We found compact radio emission from 1 supernova remnant (SN 2011dh), four AME candidates, and 8 star forming regions. We also identified several AGN which had been mis-classified as high mass X-ray binaries due to their proximity to the galaxy center. For all sources, we infer the 10GHz diffuse flux density (based on spectral indices and reported 15GHz flux densities from \cite{Linden20}, and report the ratio of 10GHz diffuse flux density to the measured 10GHz compact flux density in Tables 2 \& 3.  The full radio catalogs are described in the Appendix and available with the supplementary material.

\subsection{NGC 5474}
NGC 5474 has no resolved sources classified by \cite{Linden20}. We detect only  3 radio compact point sources.  Only one of the 3 sources matches to an X-ray source, with radio position 14:04:59.98+53:38:13 which is classified as a HMXB by \cite{2012MNRAS.419.2095M}, who identify 15 HMXB candidates total in NGC 5474.  However, optical follow-up to X-ray sources in NGC 5474 by \cite{2016MNRAS.455L..91A, 2024MNRAS.52710185A} identifies the source as a background galaxy, nearby an ultraluminous X-ray source. We do not detect radio emission from the ultraluminous X-ray source. For the background galaxy,  we measure a 10 GHz flux density of 1.97$\pm 0.04 \times 10^{-4}$ Jy, which is consistent with a low luminosity AGN.
\subsection{NGC 4631}
\label{sub:ngc4631}
After cross matching between the diffuse emission catalog from \cite{Linden20} and the compact radio catalog, we find three matches. This includes one star forming region, and two AME candidates. The positions and flux densities are displayed in Table 2. For all sources, we report a ratio between the inferred diffuse 10GHz flux density, and the measured 10GH compact flux density, to determine how much flux has been resolved out. For the AGN, we find that the measured compact flux density is considerably lower than the diffuse emission, but for the two candidate AMEs, far less flux is resolved out in the A configuration data.  

 We found one VLA source which matches to a significant ($>$ 3 $\sigma$) X-ray detection, which also matches to a region classified as SFR at position 12:42:03.59+32:32:16.28. \cite{2012MNRAS.419.2095M} classify it as a HMXB with an X-ray luminosity around $2\times 10^{37}$ erg/s, out of 26 total HMXBs identified in NGC 4631. It is about 1 arcminute from the galaxy center, and houses an AGN with jets that are approximately 5" wide  (Figure \ref{Fig:AGNs}). 
 
 Both \cite{2012MNRAS.419.2095M} and \cite{Linden20}, although they are different catalogs, working at different wavelengths with different metrics to exclude background galaxies have classified the source at position 12:42:03.59+32:32:16 as a HMXB and a SFR respectively. This demonstrates the need for multi-wavelength observations at a range of resolutions to robustly classify sources. 
                                       
\begin{table*}[]
\begin{flushleft}

\caption{M51 X-band A-configuration sources with previous LOFAR \citep{Venkattu23} and Chandra detections. Sources marked with $^l$ were detected by LOFAR. Sources marked with $^*$ were also observed by \cite{Rampadarath}. \cite{Rampadarath} classified two sources as galaxies marked with $^g$, and like \cite{Venkattu23}, we detected M51's galactic center (denoted with c), and we also detect \cite{Maddox07}'s Source 65. Three of these sources are listed in \cite{2012MNRAS.419.2095M}, which are denoted with $^m$.  $\dagger$ denotes sources identified in \cite{Maddox07}, and SNRs are denoted with s.  Only SN 2011dh was also observed by \cite{Linden20}. The positional uncertainty in R.A. and Dec is denoted as the last decimal in parentheses, and the separation between radio and X-ray is reported in arcseconds. }
\label{table:m51chandra}
\begin{tabular}{llllllllllll}
ID                    & R.A.               \& Dec.  (radio)           & X-ray Flux  & S10GHz   & Sep. \\ 
&&(0.3-10 keV) erg/s/cm$^2$& mJy & arcsec \\ \hline \hline
2CXO J132936.5+471105 & 13:29:36.58 
(3) +47:11:05.42 (5)  & 1.15 $\pm 0.43 \times 10^{-15}$         & 5.12 $\pm$ 0.81 $\times 10^{-2}$       & 0.22 \\
2CXO J132954.3+471130 & 13:29:54.31 (3) +47:11:30.07 (3)  & 1.19 $\pm 0.29 \times 10^{-15}$        & 4.71 $\pm$ 0.80 $\times 10^{-2}$       & 0.07 \\
2CXO J133016.0+471024 $^{l,*,g}$  & 13:30:16.02 (3) +47:10:24.35 (2)  & 4.51 $\pm 0.58 \times 10^{-15}$        & 7.98 $\pm$ 0.07 $\times 10^{-1}$       & 0.29 \\
2CXO J132955.8+471144 $\dagger$ $^m,s$ & 13:29:55.86 (5) +47:11:44.66 (3) & 5.70 $\pm 0.55 \times 10^{-15}$         & 3.80 $\pm$ 0.69 $\times 10^{-2}$       & 0.15 \\
2CXO J132952.6+471143 (Source 65) $\dagger$ $^l$  & 13:29:52.69 (5) +47:11:43.31 (4) & 6.55 $\pm 0.20 \times 10^{-14}$       & 4.00 $\pm$ 0.70 $\times 10^{-5}$       & 0.60 \\
2CXO J132954.9+470922 $^l,c$    & 13:29:54.96 (2) +47:09:22.69 (3) & 9.28 $\pm 0.40 \times 10^{-15}$        & 7.00  $\pm$ 0.84 $\times 10^{-2}$      & 0.15 \\
SN 2011dh$^{l,*, m,s}$  & 13:30:05.10 (4) +47:10:10.91 (3) & 1.44 $\pm 0.11 \times 10^{-14}$        & 5.07 $\pm$ 0.74 $\times 10^{-2}$       & 0.10 \\
2CXO J133011.0+471041  $\dagger$ $^{m,*,g}$ & 13:30:11.01 (2) +47:10:41.13 (1)  & 2.74 $\pm  0.12 \times10^{-14}$         & 8.79 $\pm$ 0.07 $\times 10^{-1}$       & 0.19 \\
2CXO J132959.5+471558 & 13:29:59.53 (1) +47:15:58.37 (1) & 9.29 $\pm 0.41 \times 10^{-14}$       & 3.82  $\pm$ 0.05 $\times 10^{-4}$       & 0.20 \\ \hline
\end{tabular}
\end{flushleft}

\end{table*}

\subsection{M51}    
M51 is a grand-design spiral interacting with its companion M51b actively forming stars at a rate of $\sim4.5$~M$_\odot$ yr$^{-1}$ that has been well-studied at radio and mm wavelengths \citep[e.g.,][]{2019A&A...625A..19Q}.
From cross matching between compact radio and diffuse emission, we found 10 sources matching those from \citet{Linden20} (see Table \ref{table:m51linden}). We also found a compact radio counterpart to one supernova remnant, SN2011dh, which was also observed by \cite{Rampadarath, Venkattu23}, and we also found counterparts to seven SFRs and two AME candidates from \citet{Linden20}. We again report the ratio of inferred 10GHz diffuse flux density to the measured 10GHz compact flux density, and find that very little of the flux is resolved out for SN 2011dh, as well as for several of the star forming regions. The two candidate AMEs have relatively similar flux densities in a compact array.  

From cross matching to the X-ray, we obtain nine matches between X-ray and compact radio. One of these is SN2011dh, and two are background galaxies observed by \cite{Rampadarath, Venkattu23}. Three of these match to source from \cite{2012MNRAS.419.2095M}, including a background galaxy and one SNR (also seen in \citealt{Maddox07}, source 79b).  Of the 85 HMXBs identified in M51 by \cite{2012MNRAS.419.2095M}, our radio detections rule out two contaminants.

\begin{figure}
    \centering
    \includegraphics[width=0.48\textwidth]{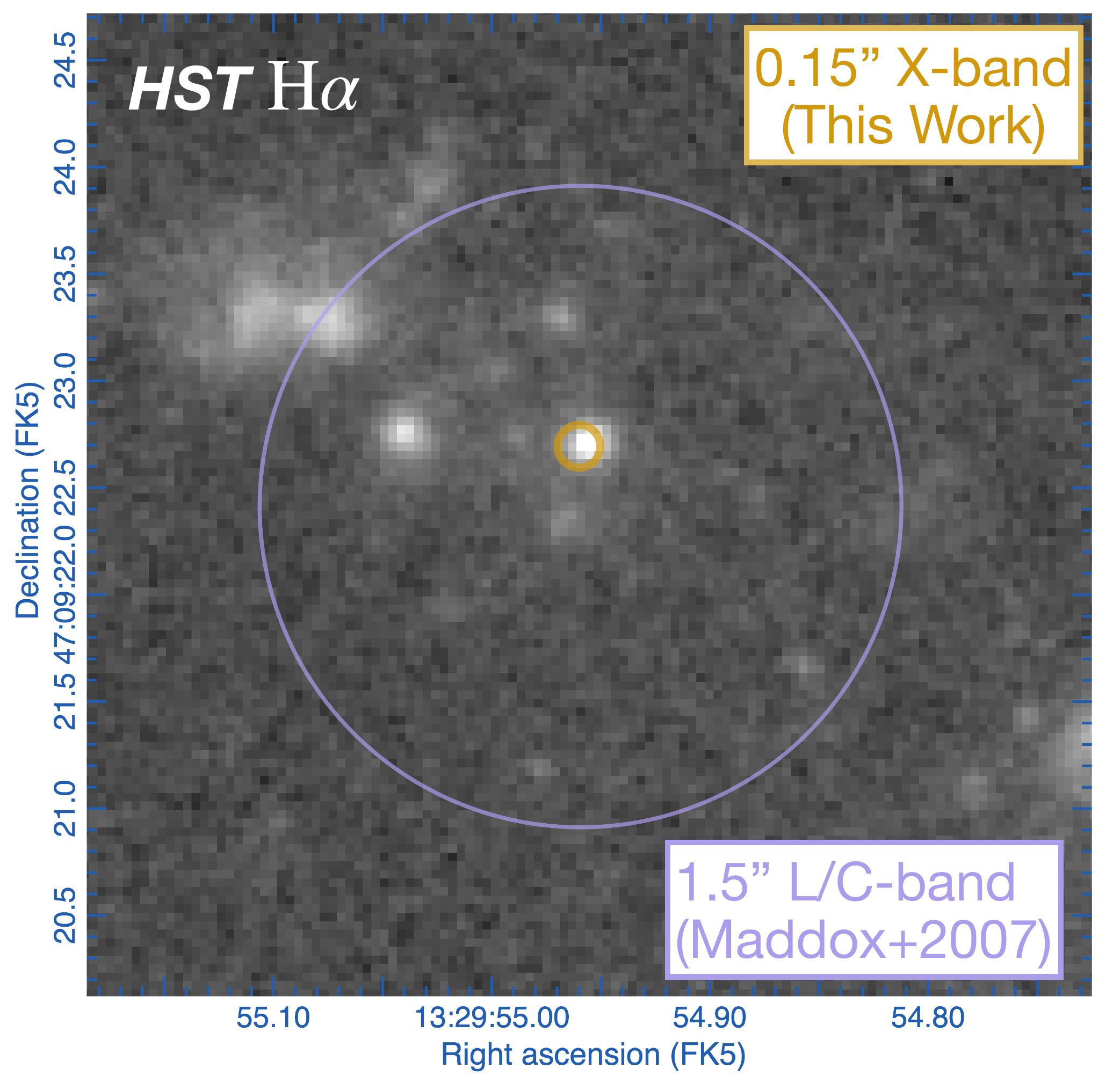}
    \caption{\emph{HST} H$\alpha$ imaging of M51 with the location of Source 65 from \citet{Maddox07} at $1.5\arcsec$ and our X-band detection ($0.15\arcsec$). The coincidence with a compact H$\alpha$ source and the steep spectrum of $\alpha=2.7\pm0.1$ suggests the source may be a radio-opaque and compact $H_{II}$ region.}
    \label{fig:m51_source65}
\end{figure}

We also detect Source 65 in \cite{Maddox07} as 2CXO J132954.9+470922 in Table \ref{table:m51chandra}, that has previous LOFAR (Source 11 in \cite{Venkattu23} and Chandra detections.
\cite{Maddox07} suggest that the X-ray source is coincident with a compact H$\alpha$ source, with a spectral index\footnote{Note that we define the spectral index $\alpha$ as $f=\nu^\alpha$.} of $\alpha$=-0.30$\pm$0.08, calculated from 20cm to 6cm. It is also listed as a HMXB candidate by \cite{2012MNRAS.419.2095M}. \cite{Venkattu23} detect it at 830 $\mu$\,Jy at 145 MHz (above 6 $\sigma$). We revisit this target at the $>10\times$ higher linear resolution relative to \citet{Maddox07} with our A-configuration observations.
We measure the flux density of the source using CARTA to fit a Gaussian source to the sub-band images (at 9 GHz and 11 GHz). From this, we find the flux density of the source at 9 GHz and 11 GHz is $45$ and $78$~$\mu$Jy, respectively. Using the flux density values at the two frequencies we were able to fit a power law and calculated the spectral index of the source to be $\alpha$=2.7$\pm$0.1. 
The steep rising spectral index we find for the compact source (Source 65) is broadly consistent with optically-thick free-free emission ($\alpha\approx2$), suggesting that the embedded cluster remains deeply embedded in what is likely---though still unresolved---a compact $H_{II}$ region.
Figure \ref{fig:m51_source65} confirms that the radio source is likely associated with an opaque $H_{II}$ region as our X-band detection associated with Source 65 also coincides with an $H_{II}$ region from archival HST narrowband H$\alpha$ imaging \citep[also used by][]{Maddox07}.
The differing spectral index we find relative to \citet{Maddox07} likely arises from diffuse free-free emission and/or other $H_{II}$ regions within their larger beam.
We also note that our estimate of the spectral index is measured over a small frequency range within X-band, which may account for the $\alpha>2$ index we measure or the presence of a non-thermal component associated with the X-ray source.

At M51's distance, our X-band resolution of $\sim0.15\arcsec$ corresponds to $\sim6$~pc scales. While this scale remains far larger than the scales of individual compact $H_{II}$ regions found in the Milky Way \citep[e.g., ][]{Churchwell2002}, the correspondence we find in Source 65 of compact H$\alpha$, opaque radio free-free and an X-ray properties consistent with a HMXB demonstrates that sensitive resolved radio surveys are entering a regime for early massive star formation that has broadly only been thoroughly studied within the Milky Way, as recent \emph{JWST} studies are beginning to show \citep{2023ApJ...944L..55L, 2023ApJ...944L..26R,2023MNRAS.520...63W, 2024ApJ...971..115G,2024ApJ...973L..55L,2024ApJ...974L..27L, Pedrini2024, 2024ApJ...967..133S, Rodriguez2025, 2025ApJ...982...50W}. Multi-band radio studies will increasingly become essential for distinguishing radio emission mechanisms within extragalactic star-forming regions.

\subsection{Diffuse and Compact Emission}

The median ratio of diffuse to compact emission for sources which overlap the catalog of \cite{Linden20} is 29.2 and 10.7 for NGC 4631 and M51 respectively. For an unresolved point source the difference in beam areas between A- and C-configuration observations at 10 GHz is ~100.  In NGC 4631 the source with the largest ratio of diffuse/compact emission is indeed an AGN that is misclassified in both \cite{2012MNRAS.419.2095M} and \cite{Linden20}. In M51, apart from SFR 7, the ratio of diffuse to compact emission is within an order of magnitude, suggesting that the majority of these sources are correctly classified as SFR or AME candidates.

\section{Summary and Conclusions} \label{sec:summary}
We present new and archival VLA A-configuration X-band analysis of three galaxies, NGC 4631, NGC 5474 and M51, and compare them to other radio and X-ray studies. Our main finding is that radio is an essential tool to diagnose contaminants in studies of extragalactic X-ray binary populations, SNe, AME, and $H_{II}$ regions hosting massive star formation, which despite multi-wavelength analyses can still be contaminated by background AGN, even within the D25 region of the host galaxy. 

We also detect seven compact radio sources likely tracing young, compact $H_{II}$ regions associated with the earliest phases of massive star formation, including likely opaque free-free emission from an $H_{II}$ region in M51.
As outlined in \cite{2024arXiv241023335P}, radio emission from star forming regions can contaminate radio and X-ray searches for intermediate mass black holes, and, moreover, sensitive resolved radio surveys further need multi-band coverage to distinguish between different radio mechanisms and opacities for star-forming regions themselves, as is often needed in Galactic studies of massive star formation \citep[e.g.,][]{Churchwell2002}.

We also identify, for the first time, compact radio sources associated with four candidate anomalous microwave emission from previous radio SED modeling by \citet{Linden20}.
This suggests that one source of AME arises from compact $H_{II}$ regions early in their evolution and further motivates IR spectroscopy with JWST to characterize these objects.

With the advent of the Square Kilometre Array \citep{2019arXiv191212699B}, and the next generation VLA \citep[][]{2018ASPC..517....3M, 2018ASPC..517...15S}, we are entering a new era with sensitive radio surveys. We advocate for the community to take full advantage of these multi-wavelength and multi-band radio surveys to further understand the role that compact radio emission plays both in understanding strange phenomena like AMEs, as well as using it to produce cleaner X-ray binary catalogs by removing background AGN and supernovae. 

This paper employs a list of Chandra datasets, obtained by the Chandra X-ray Observatory, contained in \hspace{-0.4cm} \dataset[doi:10.25574/cdc.426]{https://doi.org/10.25574/cdc.426}. All of the HST data presented in this article were obtained from the Mikulski Archive for Space Telescopes (MAST) at the Space Telescope Science Institute. The specific observations analyzed can be accessed via\dataset[doi:10.17909/x7fq-8z55]{https://doi.org/10.17909/x7fq-8z55}. 
 \begin{acknowledgments}
The authors thank the anonymous referee and the data editor for their thoughtful comments which improved the work, and thank Kat Ross for helpful discussion. KCD acknowledges support for this work provided by NASA through the NASA Hubble Fellowship grant
HST-HF2-51528 awarded by the Space Telescope Science Institute, which is operated by the Association of Universities for Research in Astronomy, Inc., for NASA, under contract NAS5–26555. 
 \end{acknowledgments}
\vspace{5mm}
\facilities{VLA, Chandra, HST}

\software{astropy \citep{Robitaille13}, CASA \citep{CASA2022}, matplotlib \citep{Hunter07}, NumPy \citep{harris20}, pandas \citep{Mckinney10}, PyBDSF \citep{2015ascl.soft02007M}}

\bibliography{draft}{}

\begin{thebibliography}{}
\expandafter\ifx\csname natexlab\endcsname\relax\def\natexlab#1{#1}\fi
\providecommand{\url}[1]{\href{#1}{#1}}
\providecommand{\dodoi}[1]{doi:~\href{http://doi.org/#1}{\nolinkurl{#1}}}
\providecommand{\doeprint}[1]{\href{http://ascl.net/#1}{\nolinkurl{http://ascl.net/#1}}}
\providecommand{\doarXiv}[1]{\href{https://arxiv.org/abs/#1}{\nolinkurl{https://arxiv.org/abs/#1}}}

\bibitem[{{Astropy Collaboration} {et~al.}(2013){Astropy Collaboration}, {Robitaille}, {Tollerud}, {Greenfield}, {Droettboom}, {Bray}, {Aldcroft}, {Davis}, {Ginsburg}, {Price-Whelan}, {Kerzendorf}, {Conley}, {Crighton}, {Barbary}, {Muna}, {Ferguson}, {Grollier}, {Parikh}, {Nair}, {Unther}, {Deil}, {Woillez}, {Conseil}, {Kramer}, {Turner}, {Singer}, {Fox}, {Weaver}, {Zabalza}, {Edwards}, {Azalee Bostroem}, {Burke}, {Casey}, {Crawford}, {Dencheva}, {Ely}, {Jenness}, {Labrie}, {Lim}, {Pierfederici}, {Pontzen}, {Ptak}, {Refsdal}, {Servillat}, \& {Streicher}}]{Robitaille13}
{Astropy Collaboration}, {Robitaille}, T.~P., {Tollerud}, E.~J., {et~al.} 2013, \aap, 558, A33, \dodoi{10.1051/0004-6361/201322068}

\bibitem[{{Atapin} {et~al.}(2024){Atapin}, {Vinokurov}, {Sarkisyan}, {Solovyeva}, {Kostenkov}, {Medvedev}, \& {Fabrika}}]{2024MNRAS.52710185A}
{Atapin}, K., {Vinokurov}, A., {Sarkisyan}, A., {et~al.} 2024, \mnras, 527, 10185, \dodoi{10.1093/mnras/stad3719}

\bibitem[{{Avdan} {et~al.}(2016){Avdan}, {Vinokurov}, {Fabrika}, {Atapin}, {Avdan}, {Akyuz}, {Sholukhova}, {Aksaker}, \& {Valeev}}]{2016MNRAS.455L..91A}
{Avdan}, S., {Vinokurov}, A., {Fabrika}, S., {et~al.} 2016, \mnras, 455, L91, \dodoi{10.1093/mnrasl/slv155}

\bibitem[{{Beck}(2015)}]{Beck2015}
{Beck}, R. 2015, \aapr, 24, 4, \dodoi{10.1007/s00159-015-0084-4}

\bibitem[{{Braun} {et~al.}(2019){Braun}, {Bonaldi}, {Bourke}, {Keane}, \& {Wagg}}]{2019arXiv191212699B}
{Braun}, R., {Bonaldi}, A., {Bourke}, T., {Keane}, E., \& {Wagg}, J. 2019, arXiv e-prints, arXiv:1912.12699, \dodoi{10.48550/arXiv.1912.12699}

\bibitem[{{Briggs}(1995)}]{1995briggs}
{Briggs}, D.~S. 1995, in American Astronomical Society Meeting Abstracts, Vol. 187, American Astronomical Society Meeting Abstracts, 112.02

\bibitem[{{Calzetti} {et~al.}(2015){Calzetti}, {Lee}, {Sabbi}, {Adamo}, {Smith}, {Andrews}, {Ubeda}, {Bright}, {Thilker}, {Aloisi}, {Brown}, {Chandar}, {Christian}, {Cignoni}, {Clayton}, {da Silva}, {de Mink}, {Dobbs}, {Elmegreen}, {Elmegreen}, {Evans}, {Fumagalli}, {Gallagher}, {Gouliermis}, {Grebel}, {Herrero}, {Hunter}, {Johnson}, {Kennicutt}, {Kim}, {Krumholz}, {Lennon}, {Levay}, {Martin}, {Nair}, {Nota}, {{\"O}stlin}, {Pellerin}, {Prieto}, {Regan}, {Ryon}, {Schaerer}, {Schiminovich}, {Tosi}, {Van Dyk}, {Walterbos}, {Whitmore}, \& {Wofford}}]{2015AJ....149...51C}
{Calzetti}, D., {Lee}, J.~C., {Sabbi}, E., {et~al.} 2015, \aj, 149, 51, \dodoi{10.1088/0004-6256/149/2/51}

\bibitem[{{CASA Team} {et~al.}(2022{\natexlab{a}}){CASA Team}, {Bean}, {Bhatnagar}, {Castro}, {Donovan Meyer}, {Emonts}, {Garcia}, {Garwood}, {Golap}, {Gonzalez Villalba}, {Harris}, {Hayashi}, {Hoskins}, {Hsieh}, {Jagannathan}, {Kawasaki}, {Keimpema}, {Kettenis}, {Lopez}, {Marvil}, {Masters}, {McNichols}, {Mehringer}, {Miel}, {Moellenbrock}, {Montesino}, {Nakazato}, {Ott}, {Petry}, {Pokorny}, {Raba}, {Rau}, {Schiebel}, {Schweighart}, {Sekhar}, {Shimada}, {Small}, {Steeb}, {Sugimoto}, {Suoranta}, {Tsutsumi}, {van Bemmel}, {Verkouter}, {Wells}, {Xiong}, {Szomoru}, {Griffith}, {Glendenning}, \& {Kern}}]{2022casa}
{CASA Team}, {Bean}, B., {Bhatnagar}, S., {et~al.} 2022{\natexlab{a}}, \pasp, 134, 114501, \dodoi{10.1088/1538-3873/ac9642}

\bibitem[{{CASA Team} {et~al.}(2022{\natexlab{b}}){CASA Team}, {Bean}, {Bhatnagar}, {Castro}, {Donovan Meyer}, {Emonts}, {Garcia}, {Garwood}, {Golap}, {Gonzalez Villalba}, {Harris}, {Hayashi}, {Hoskins}, {Hsieh}, {Jagannathan}, {Kawasaki}, {Keimpema}, {Kettenis}, {Lopez}, {Marvil}, {Masters}, {McNichols}, {Mehringer}, {Miel}, {Moellenbrock}, {Montesino}, {Nakazato}, {Ott}, {Petry}, {Pokorny}, {Raba}, {Rau}, {Schiebel}, {Schweighart}, {Sekhar}, {Shimada}, {Small}, {Steeb}, {Sugimoto}, {Suoranta}, {Tsutsumi}, {van Bemmel}, {Verkouter}, {Wells}, {Xiong}, {Szomoru}, {Griffith}, {Glendenning}, \& {Kern}}]{CASA2022}
---. 2022{\natexlab{b}}, \pasp, 134, 114501, \dodoi{10.1088/1538-3873/ac9642}

\bibitem[{{Churchwell}(2002)}]{Churchwell2002}
{Churchwell}, E. 2002, \araa, 40, 27, \dodoi{10.1146/annurev.astro.40.060401.093845}

\bibitem[{{Condon}(1987)}]{Condon1987}
{Condon}, J.~J. 1987, \apjs, 65, 485, \dodoi{10.1086/191234}

\bibitem[{{de Oliveira-Costa} {et~al.}(1997){de Oliveira-Costa}, {Kogut}, {Devlin}, {Netterfield}, {Page}, \& {Wollack}}]{deOliveira-Costa1997}
{de Oliveira-Costa}, A., {Kogut}, A., {Devlin}, M.~J., {et~al.} 1997, \apjl, 482, L17, \dodoi{10.1086/310684}

\bibitem[{{Draine} \& {Lazarian}(1998{\natexlab{a}})}]{Draine1998a}
{Draine}, B.~T., \& {Lazarian}, A. 1998{\natexlab{a}}, \apjl, 494, L19, \dodoi{10.1086/311167}

\bibitem[{{Draine} \& {Lazarian}(1998{\natexlab{b}})}]{Draine1998b}
---. 1998{\natexlab{b}}, \apj, 508, 157, \dodoi{10.1086/306387}

\bibitem[{{Draine} \& {Lazarian}(1999)}]{Draine1999}
---. 1999, \apj, 512, 740, \dodoi{10.1086/306809}

\bibitem[{{Fruscione} {et~al.}(2006){Fruscione}, {McDowell}, {Allen}, {Brickhouse}, {Burke}, {Davis}, {Durham}, {Elvis}, {Galle}, {Harris}, {Huenemoerder}, {Houck}, {Ishibashi}, {Karovska}, {Nicastro}, {Noble}, {Nowak}, {Primini}, {Siemiginowska}, {Smith}, \& {Wise}}]{Fruscione06}
{Fruscione}, A., {McDowell}, J.~C., {Allen}, G.~E., {et~al.} 2006, in Society of Photo-Optical Instrumentation Engineers (SPIE) Conference Series, Vol. 6270, 62701V, \dodoi{10.1117/12.671760}

\bibitem[{{Gregg} {et~al.}(2024){Gregg}, {Calzetti}, {Adamo}, {Bajaj}, {Ryon}, {Linden}, {Correnti}, {Cignoni}, {Messa}, {Sabbi}, {Gallagher}, {Grasha}, {Pedrini}, {Gutermuth}, {Melinder}, {Kotulla}, {P{\'e}rez}, {Krumholz}, {Bik}, {{\"O}stlin}, {Johnson}, {Bortolini}, {Smith}, {Tosi}, {Maji}, \& {Faustino Vieira}}]{2024ApJ...971..115G}
{Gregg}, B., {Calzetti}, D., {Adamo}, A., {et~al.} 2024, \apj, 971, 115, \dodoi{10.3847/1538-4357/ad54b4}

\bibitem[{Harris {et~al.}(2020)Harris, Millman, van~der Walt, Gommers, Virtanen, Cournapeau, Wieser, Taylor, Berg, Smith, Kern, Picus, Hoyer, van Kerkwijk, Brett, Haldane, del R{'{\i}}o, Wiebe, Peterson, G{'{e}}rard-Marchant, Sheppard, Reddy, Weckesser, Abbasi, Gohlke, \& Oliphant}]{harris20}
Harris, C.~R., Millman, K.~J., van~der Walt, S.~J., {et~al.} 2020, Nature, 585, 357, \dodoi{10.1038/s41586-020-2649-2}

\bibitem[{Hunter(2007)}]{Hunter07}
Hunter, J.~D. 2007, Computing In Science \& Engineering, 9, 90, \dodoi{10.1109/MCSE.2007.55}

\bibitem[{{Kogut} {et~al.}(1996){Kogut}, {Banday}, {Bennett}, {Gorski}, {Hinshaw}, \& {Reach}}]{Kogut1996}
{Kogut}, A., {Banday}, A.~J., {Bennett}, C.~L., {et~al.} 1996, \apj, 460, 1, \dodoi{10.1086/176947}

\bibitem[{{Leitch} {et~al.}(1997){Leitch}, {Readhead}, {Pearson}, \& {Myers}}]{Leitch1997}
{Leitch}, E.~M., {Readhead}, A.~C.~S., {Pearson}, T.~J., \& {Myers}, S.~T. 1997, \apjl, 486, L23, \dodoi{10.1086/310823}

\bibitem[{{Leroy} {et~al.}(2019){Leroy}, {Sandstrom}, {Lang}, {Lewis}, {Salim}, {Behrens}, {Chastenet}, {Chiang}, {Gallagher}, {Kessler}, \& {Utomo}}]{Leroy2019}
{Leroy}, A.~K., {Sandstrom}, K.~M., {Lang}, D., {et~al.} 2019, \apjs, 244, 24, \dodoi{10.3847/1538-4365/ab3925}

\bibitem[{{Levy} {et~al.}(2024){Levy}, {Bolatto}, {Mayya}, {Cuevas-Otahola}, {Tarantino}, {Boyer}, {Boogaard}, {B{\"o}ker}, {Cronin}, {Dale}, {Donaghue}, {Emig}, {Fisher}, {Glover}, {Herrera-Camus}, {Jim{\'e}nez-Donaire}, {Klessen}, {Lenki{\'c}}, {Leroy}, {De Looze}, {Meier}, {Mills}, {Ott}, {Rela{\~n}o}, {Veilleux}, {Villanueva}, {Walter}, \& {van der Werf}}]{2024ApJ...973L..55L}
{Levy}, R.~C., {Bolatto}, A.~D., {Mayya}, D., {et~al.} 2024, \apjl, 973, L55, \dodoi{10.3847/2041-8213/ad7af3}

\bibitem[{{Linden} \& {Mihos}(2022)}]{Linden2022}
{Linden}, S.~T., \& {Mihos}, J.~C. 2022, \apjl, 933, L33, \dodoi{10.3847/2041-8213/ac7c06}

\bibitem[{{Linden} {et~al.}(2020){Linden}, {Murphy}, {Dong}, {Momjian}, {Kennicutt}, {Meier}, {Schinnerer}, \& {Turner}}]{Linden20}
{Linden}, S.~T., {Murphy}, E.~J., {Dong}, D., {et~al.} 2020, \apjs, 248, 25, \dodoi{10.3847/1538-4365/ab8a4d}

\bibitem[{{Linden} {et~al.}(2023){Linden}, {Evans}, {Armus}, {Rich}, {Larson}, {Lai}, {Privon}, {U}, {Inami}, {Bohn}, {Song}, {Barcos-Mu{\~n}oz}, {Charmandaris}, {Medling}, {Stierwalt}, {Diaz-Santos}, {B{\"o}ker}, {van der Werf}, {Aalto}, {Appleton}, {Brown}, {Hayward}, {Howell}, {Iwasawa}, {Kemper}, {Frayer}, {Law}, {Malkan}, {Marshall}, {Mazzarella}, {Murphy}, {Sanders}, \& {Surace}}]{2023ApJ...944L..55L}
{Linden}, S.~T., {Evans}, A.~S., {Armus}, L., {et~al.} 2023, \apjl, 944, L55, \dodoi{10.3847/2041-8213/acb335}

\bibitem[{{Linden} {et~al.}(2024){Linden}, {Lai}, {Evans}, {Armus}, {Larson}, {Rich}, {U}, {Privon}, {Inami}, {Song}, {Bianchin}, {Bohn}, {Buiten}, {Sanchez-Garc{\'\i}a}, {Kader}, {Lenki{\'c}}, {Medling}, {B{\"o}ker}, {D{\'\i}az-Santos}, {Charmandaris}, {Barcos-Mu{\~n}oz}, {van der Werf}, {Stierwalt}, {Aalto}, {Appleton}, {Hayward}, {Howell}, {Malkan}, {Mazzarella}, {Murphy}, \& {Surace}}]{2024ApJ...974L..27L}
{Linden}, S.~T., {Lai}, T., {Evans}, A.~S., {et~al.} 2024, \apjl, 974, L27, \dodoi{10.3847/2041-8213/ad7eae}

\bibitem[{{Maddox} {et~al.}(2007){Maddox}, {Cowan}, {Kilgard}, {Schinnerer}, \& {Stockdale}}]{Maddox07}
{Maddox}, L.~A., {Cowan}, J.~J., {Kilgard}, R.~E., {Schinnerer}, E., \& {Stockdale}, C.~J. 2007, \aj, 133, 2559, \dodoi{10.1086/515573}

\bibitem[{McKinney(2010)}]{Mckinney10}
McKinney, W. 2010, in Proceedings of the 9th Python in Science Conference, ed. S.~van~der Walt \& J.~Millman, 51 -- 56

\bibitem[{{McMullin} {et~al.}(2007){McMullin}, {Waters}, {Schiebel}, {Young}, \& {Golap}}]{2007casa}
{McMullin}, J.~P., {Waters}, B., {Schiebel}, D., {Young}, W., \& {Golap}, K. 2007, in Astronomical Society of the Pacific Conference Series, Vol. 376, Astronomical Data Analysis Software and Systems XVI, ed. R.~A. {Shaw}, F.~{Hill}, \& D.~J. {Bell}, 127

\bibitem[{{McQuinn} {et~al.}(2016){McQuinn}, {Skillman}, {Dolphin}, {Berg}, \& {Kennicutt}}]{McQuinn16}
{McQuinn}, K. B.~W., {Skillman}, E.~D., {Dolphin}, A.~E., {Berg}, D., \& {Kennicutt}, R. 2016, \apj, 826, 21, \dodoi{10.3847/0004-637X/826/1/21}

\bibitem[{{Mineo} {et~al.}(2012){Mineo}, {Gilfanov}, \& {Sunyaev}}]{2012MNRAS.419.2095M}
{Mineo}, S., {Gilfanov}, M., \& {Sunyaev}, R. 2012, \mnras, 419, 2095, \dodoi{10.1111/j.1365-2966.2011.19862.x}

\bibitem[{{Mohan} \& {Rafferty}(2015)}]{2015ascl.soft02007M}
{Mohan}, N., \& {Rafferty}, D. 2015, {PyBDSF: Python Blob Detection and Source Finder}, Astrophysics Source Code Library, record ascl:1502.007

\bibitem[{{Mora} \& {Krause}(2013)}]{Mora2013}
{Mora}, S.~C., \& {Krause}, M. 2013, \aap, 560, A42, \dodoi{10.1051/0004-6361/201321043}

\bibitem[{{Murphy} {et~al.}(2010){Murphy}, {Helou}, {Condon}, {Schinnerer}, {Turner}, {Beck}, {Mason}, {Chary}, \& {Armus}}]{Murphy2010}
{Murphy}, E.~J., {Helou}, G., {Condon}, J.~J., {et~al.} 2010, \apjl, 709, L108, \dodoi{10.1088/2041-8205/709/2/L108}

\bibitem[{{Murphy} {et~al.}(2018){Murphy}, {Bolatto}, {Chatterjee}, {Casey}, {Chomiuk}, {Dale}, {de Pater}, {Dickinson}, {Francesco}, {Hallinan}, {Isella}, {Kohno}, {Kulkarni}, {Lang}, {Lazio}, {Leroy}, {Loinard}, {Maccarone}, {Matthews}, {Osten}, {Reid}, {Riechers}, {Sakai}, {Walter}, \& {Wilner}}]{2018ASPC..517....3M}
{Murphy}, E.~J., {Bolatto}, A., {Chatterjee}, S., {et~al.} 2018, in Astronomical Society of the Pacific Conference Series, Vol. 517, Science with a Next Generation Very Large Array, ed. E.~{Murphy}, 3, \dodoi{10.48550/arXiv.1810.07524}

\bibitem[{{Panurach} {et~al.}(2024){Panurach}, {Dage}, {Urquhart}, {Plotkin}, {Paul}, {Bahramian}, {Brumback}, {Galvin}, {Molina}, {Miller-Jones}, \& {Saikia}}]{2024arXiv241023335P}
{Panurach}, T., {Dage}, K.~C., {Urquhart}, R., {et~al.} 2024, arXiv e-prints, arXiv:2410.23335, \dodoi{10.48550/arXiv.2410.23335}

\bibitem[{{Paturel} {et~al.}(1987){Paturel}, {Fouque}, {Lauberts}, {Valentijn}, {Corwin}, \& {de Vaucouleurs}}]{Paturel1987}
{Paturel}, G., {Fouque}, P., {Lauberts}, A., {et~al.} 1987, \aap, 184, 86

\bibitem[{{Pedrini} {et~al.}(2024){Pedrini}, {Adamo}, {Calzetti}, {Bik}, {Gregg}, {Linden}, {Bajaj}, {Ryon}, {Ali}, {Bortolini}, {Correnti}, {Elmegreen}, {Elmegreen}, {Gallagher}, {Grasha}, {Gutermuth}, {Johnson}, {Melinder}, {Messa}, {{\"O}stlin}, {Sabbi}, {Smith}, {Tosi}, \& {Faustino Vieira}}]{Pedrini2024}
{Pedrini}, A., {Adamo}, A., {Calzetti}, D., {et~al.} 2024, \apj, 971, 32, \dodoi{10.3847/1538-4357/ad534d}

\bibitem[{{Querejeta} {et~al.}(2019){Querejeta}, {Schinnerer}, {Schruba}, {Murphy}, {Meidt}, {Usero}, {Leroy}, {Pety}, {Bigiel}, {Chevance}, {Faesi}, {Gallagher}, {Garc{\'\i}a-Burillo}, {Glover}, {Hygate}, {Jim{\'e}nez-Donaire}, {Kruijssen}, {Momjian}, {Rosolowsky}, \& {Utomo}}]{2019A&A...625A..19Q}
{Querejeta}, M., {Schinnerer}, E., {Schruba}, A., {et~al.} 2019, \aap, 625, A19, \dodoi{10.1051/0004-6361/201834915}

\bibitem[{{Rampadarath} {et~al.}(2015){Rampadarath}, {Morgan}, {Soria}, {Tingay}, {Reynolds}, {Argo}, \& {Dumas}}]{Rampadarath}
{Rampadarath}, H., {Morgan}, J.~S., {Soria}, R., {et~al.} 2015, \mnras, 452, 32, \dodoi{10.1093/mnras/stv1275}

\bibitem[{{Rau}(2012)}]{2012rau}
{Rau}, U. 2012, in Society of Photo-Optical Instrumentation Engineers (SPIE) Conference Series, Vol. 8500, Image Reconstruction from Incomplete Data VII, ed. P.~J. {Bones}, M.~A. {Fiddy}, \& R.~P. {Millane}, 85000N, \dodoi{10.1117/12.930207}

\bibitem[{{Rodr{\'\i}guez} {et~al.}(2023){Rodr{\'\i}guez}, {Lee}, {Whitmore}, {Thilker}, {Maschmann}, {Chandar}, {Deger}, {Boquien}, {Dale}, {Larson}, {Williams}, {Kim}, {Schinnerer}, {Rosolowsky}, {Leroy}, {Emsellem}, {Sandstrom}, {Kruijssen}, {Grasha}, {Watkins}, {Barnes}, {Sormani}, {Kim}, {Anand}, {Chevance}, {Bigiel}, {Klessen}, {Hassani}, {Liu}, {Faesi}, {Cao}, {Belfiore}, {Pessa}, {Kreckel}, {Groves}, {Pety}, {Indebetouw}, {Egorov}, {Blanc}, {Saito}, \& {Hughes}}]{2023ApJ...944L..26R}
{Rodr{\'\i}guez}, M.~J., {Lee}, J.~C., {Whitmore}, B.~C., {et~al.} 2023, \apjl, 944, L26, \dodoi{10.3847/2041-8213/aca653}

\bibitem[{{Rodr{\'\i}guez} {et~al.}(2025){Rodr{\'\i}guez}, {Lee}, {Indebetouw}, {Whitmore}, {Maschmann}, {Williams}, {Chandar}, {Barnes}, {Gnedin}, {Sandstrom}, {Rosolowsky}, {Leroy}, {Thilker}, {Kim}, {Sun}, {Klessen}, {Groves}, {Wofford}, {Boquien}, {Dale}, {{\'U}beda}, {Larson}, {Grasha}, {Johnson}, {Levy}, {Bigiel}, {Hassani}, \& {Sarbadhicary}}]{Rodriguez2025}
{Rodr{\'\i}guez}, M.~J., {Lee}, J.~C., {Indebetouw}, R., {et~al.} 2025, \apj, 983, 137, \dodoi{10.3847/1538-4357/adbb69}

\bibitem[{{Selina} {et~al.}(2018){Selina}, {Murphy}, {McKinnon}, {Beasley}, {Butler}, {Carilli}, {Clark}, {Durand}, {Erickson}, {Grammer}, {Hiriart}, {Jackson}, {Kent}, {Mason}, {Morgan}, {Ojeda}, {Rosero}, {Shillue}, {Sturgis}, \& {Urbain}}]{2018ASPC..517...15S}
{Selina}, R.~J., {Murphy}, E.~J., {McKinnon}, M., {et~al.} 2018, in Astronomical Society of the Pacific Conference Series, Vol. 517, Science with a Next Generation Very Large Array, ed. E.~{Murphy}, 15, \dodoi{10.48550/arXiv.1810.08197}

\bibitem[{{Sun} {et~al.}(2024){Sun}, {He}, {Batschkun}, {Levy}, {Emig}, {Rodr{\'\i}guez}, {Hassani}, {Leroy}, {Schinnerer}, {Ostriker}, {Wilson}, {Bolatto}, {Mills}, {Rosolowsky}, {Lee}, {Dale}, {Larson}, {Thilker}, {Ubeda}, {Whitmore}, {Williams}, {Barnes}, {Bigiel}, {Chevance}, {Glover}, {Grasha}, {Groves}, {Henshaw}, {Indebetouw}, {Jim{\'e}nez-Donaire}, {Klessen}, {Koch}, {Liu}, {Mathur}, {Meidt}, {Menon}, {Neumann}, {Pinna}, {Querejeta}, {Sormani}, \& {Tress}}]{2024ApJ...967..133S}
{Sun}, J., {He}, H., {Batschkun}, K., {et~al.} 2024, \apj, 967, 133, \dodoi{10.3847/1538-4357/ad3de6}

\bibitem[{{Thompson} {et~al.}(1980){Thompson}, {Clark}, {Wade}, \& {Napier}}]{1980ApJS...44..151T}
{Thompson}, A.~R., {Clark}, B.~G., {Wade}, C.~M., \& {Napier}, P.~J. 1980, \apjs, 44, 151, \dodoi{10.1086/190688}

\bibitem[{{Tikhonov} {et~al.}(2015){Tikhonov}, {Lebedev}, \& {Galazutdinova}}]{2015AstL...41..239T}
{Tikhonov}, N.~A., {Lebedev}, V.~S., \& {Galazutdinova}, O.~A. 2015, Astronomy Letters, 41, 239, \dodoi{10.1134/S1063773715060080}

\bibitem[{{Tully} {et~al.}(2013){Tully}, {Courtois}, {Dolphin}, {Fisher}, {H{\'e}raudeau}, {Jacobs}, {Karachentsev}, {Makarov}, {Makarova}, {Mitronova}, {Rizzi}, {Shaya}, {Sorce}, \& {Wu}}]{2013AJ....146...86T}
{Tully}, R.~B., {Courtois}, H.~M., {Dolphin}, A.~E., {et~al.} 2013, \aj, 146, 86, \dodoi{10.1088/0004-6256/146/4/86}

\bibitem[{{Venkattu} {et~al.}(2023){Venkattu}, {Lundqvist}, {P{\'e}rez Torres}, {Morabito}, {Mold{\'o}n}, {Conway}, {Chandra}, \& {Tasse}}]{Venkattu23}
{Venkattu}, D., {Lundqvist}, P., {P{\'e}rez Torres}, M., {et~al.} 2023, \apj, 953, 157, \dodoi{10.3847/1538-4357/ace2c1}

\bibitem[{{Wang} {et~al.}(1995){Wang}, {Walterbos}, {Steakley}, {Norman}, \& {Braun}}]{Wang1995}
{Wang}, Q.~D., {Walterbos}, R. A.~M., {Steakley}, M.~F., {Norman}, C.~A., \& {Braun}, R. 1995, \apj, 439, 176, \dodoi{10.1086/175162}

\bibitem[{{Whitmore} {et~al.}(2023){Whitmore}, {Chandar}, {Lee}, {Floyd}, {Deger}, {Lilly}, {Minsley}, {Thilker}, {Boquien}, {Dale}, {Henny}, {Scheuermann}, {Barnes}, {Bigiel}, {Emsellem}, {Glover}, {Grasha}, {Groves}, {Hannon}, {Klessen}, {Kreckel}, {Kruijssen}, {Larson}, {Leroy}, {Mok}, {Pan}, {Pinna}, {S{\'a}nchez-Bl{\'a}zquez}, {Schinnerer}, {Sormani}, {Watkins}, \& {Williams}}]{2023MNRAS.520...63W}
{Whitmore}, B.~C., {Chandar}, R., {Lee}, J.~C., {et~al.} 2023, \mnras, 520, 63, \dodoi{10.1093/mnras/stad098}

\bibitem[{{Whitmore} {et~al.}(2025){Whitmore}, {Chandar}, {Lee}, {Henny}, {Rodr{\'\i}guez}, {Baron}, {Bigiel}, {Boquien}, {Chevance}, {Chown}, {Dale}, {Floyd}, {Grasha}, {Glover}, {Gnedin}, {Hassani}, {Indebetouw}, {Kapoor}, {Larson}, {Leroy}, {Maschmann}, {Scheuermann}, {Sutter}, {Schinnerer}, {Sarbadhicary}, {Thilker}, {Williams}, \& {Wofford}}]{2025ApJ...982...50W}
---. 2025, \apj, 982, 50, \dodoi{10.3847/1538-4357/adb3a2}

\bibitem[{{Wiegert} {et~al.}(2015){Wiegert}, {Irwin}, {Miskolczi}, {Schmidt}, {Mora}, {Damas-Segovia}, {Stein}, {English}, {Rand}, {Santistevan}, {Walterbos}, {Krause}, {Beck}, {Dettmar}, {Kepley}, {Wezgowiec}, {Wang}, {Heald}, {Li}, {MacGregor}, {Johnson}, {Strong}, {DeSouza}, \& {Porter}}]{Wiegert2015}
{Wiegert}, T., {Irwin}, J., {Miskolczi}, A., {et~al.} 2015, \aj, 150, 81, \dodoi{10.1088/0004-6256/150/3/81}

\bibitem[{{Yu} {et~al.}(2023){Yu}, {Zhu}, {Xu}, {Ai}, {Jiang}, \& {Yang}}]{2023MNRAS.521.2719Y}
{Yu}, H., {Zhu}, M., {Xu}, J.-L., {et~al.} 2023, \mnras, 521, 2719, \dodoi{10.1093/mnras/stad436}

\end{thebibliography}
\bibliographystyle{aasjournal}

\appendix

\section{Complete X-band source catalogs}
\label{app:fullcatalogs}

This paper focuses only on X-band sources that have existing multi-band radio or X-ray detections.
For completeness, we include supplementary catalogs with all X-band radio sources identified using PyBDSF (see \S\ref{sec:results}).
In total, we detect 66 sources in NGC~5194, 46 in NGC~4631, and 3 in NGC~5474 above with $>5\sigma$ significance, with the exception of a handful that are detected between 4 and 5 $\sigma$. These catalogs are available in the supplementary material. A snippet of the supplemental material is presented in Table 5, showing the top ten rows of the most relevant columns generated by pyBDSF. The full catalogs include this, along with total flux \& total flux error (Total\_flux, E\_Total\_flux), the R.A. and Dec of the maximum of the source (RA\_max, DEC\_max) and their corresponding errors (E\_RA\_max, E\_DEC\_max; in degrees), the FWHM of the major and minor axis (Maj, Min) and corresponding errors (E\_Maj, E\_Min; in degrees), the deconvolved major and minor axis of the source and corresponding errors (DC\_Maj, DC\_Min, E\_DC\_Maj, E\_DC\_Min) the position angle of the source and corresponding error (PA, E\_PA, DC\_PA, E\_DC\_PA), and similar information for the source in the image plane (Maj\_img\_plane, E\_Maj\_img\_plane, Min\_img\_plane, E\_Min\_img\_plane, DC\_Maj\_img\_plane, E\_DC\_Maj\_img\_plane, DC\_Min\_img\_plane, E\_DC\_Min\_img\_plane), and the island total flux density, error, average background RMS, mean \& residuals (Isl\_Total\_Flux, E\_Isl\_Total\_Flux, Isl\_rms, Isl\_mean, Resid\_Isl\_rms, Resid\_Isl\_mean, source code (denoting island structure; S\_code). The pybdsf documentation provide a full description\footnote{\url{https://pybdsf.readthedocs.io/en/latest/write_catalog.html}}.
\begin{table}[h]
\label{table:appendix}
    \centering
    \caption{Selected point source catalog of NGC 4631, available in the supplemental information, including source ID, R.A., R.A. error, Dec., Dec. error, and the peak flux density in Janskys in a point source (used for analysis in this paper) and the corresponding error.}

\begin{tabular}{lllllll}
ID & RA                 & E\_RA         & DEC               & E\_DEC        & Peak\_flux     & E\_Peak\_flux \\
0  & 190.55949892243098 & 2.60037291E-6 & 32.56602021909703 & 2.5562414E-6  & 2.255560021E-5 & 3.94780742E-6 \\
1  & 190.55473289402894 & 2.91920184E-6 & 32.562948695478   & 2.42118541E-6 & 2.06779542E-5  & 3.81461081E-6 \\
2  & 190.5538142454355  & 2.70403793E-6 & 32.5362418284849  & 3.46259851E-6 & 2.131554885E-5 & 4.07299274E-6 \\
3  & 190.55370767315264 & 3.12947631E-6 & 32.54310034123638 & 2.28523246E-6 & 3.23664802E-5  & 4.35550609E-6 \\
4  & 190.55081403941836 & 2.84684281E-6 & 32.52138538451722 & 2.12146306E-6 & 2.245889763E-5 & 3.8359067E-6  \\
5  & 190.5485929657391  & 2.57252195E-6 & 32.54025472514552 & 3.34265993E-6 & 2.105883552E-5 & 3.91784664E-6 \\
6  & 190.5458812415038  & 2.82726422E-6 & 32.54381724648358 & 2.75939067E-6 & 3.821724063E-5 & 4.44585957E-6 \\
7  & 190.545801027826   & 2.00347111E-6 & 32.54275794774701 & 1.51729708E-6 & 5.265356452E-5 & 4.49204144E-6 \\
8  & 190.5452865859493  & 4.12768456E-6 & 32.54303861029288 & 4.885667E-6   & 2.220012131E-5 & 4.42590777E-6 \\
9  & 190.5451318154608  & 4.28556091E-6 & 32.54302262643947 & 2.49438515E-6 & 2.128504252E-5 & 4.11811525E-6
\end{tabular}
\end{table}

\end{document}